\documentclass[10pt,final,journal]{IEEEtran}
\IEEEoverridecommandlockouts 
\usepackage{amsmath,amssymb,amsfonts}
\usepackage{graphicx}
\usepackage{cite}
\usepackage[keeplastbox]{flushend} 
\usepackage{tikz}
\usepackage{balance}
\usepackage{caption}
\usepackage{epstopdf} 
\usetikzlibrary{automata, positioning}
\hfuzz 
\vfuzz
\hbadness 
\vbadness
\newcommand{\RN}[1]{%
\textup{\uppercase\expandafter{\romannumeral#1}}
}
\begin{document}
\title{HAPS Selection for Hybrid RF/FSO Satellite Networks}
\author{Olfa Ben Yahia, Eylem Erdogan, \textit{Senior Member}, \textit{IEEE}, Gunes~Karabulut~Kurt, \textit{Senior Member}, \textit{IEEE}, \\ 
Ibrahim Altunbas, \textit{Senior Member}, \textit{IEEE}, Halim Yanikomeroglu, \textit{Fellow}, \textit{IEEE}%
\thanks{O. Ben Yahia, G. ~Karabulut~Kurt and I. Altunbas are with the Department of Electronics and Communication Engineering, Istanbul Technical University, Istanbul, Turkey, (e-mails: \{yahiao17, gkurt, ibraltunbas\}@itu.edu.tr).}%
\thanks{E. Erdogan is with the Department of Electrical and Electronics Engineering, Istanbul Medeniyet University, Istanbul, Turkey, (e-mail: eylem.erdogan@medeniyet.edu.tr). }%
\thanks{G.~Karabulut~Kurt is also with the Department of Electrical Engineering, Polytechnique Montréal, Montréal, QC, Canada (e-mail: gunes.kurt@polymtl.ca). }%
\thanks{H. Yanikomeroglu is with the Department of Systems and Computer Engineering, Carleton University, Ottawa, ON, Canada, (e-mail: halim@sce.carleton.ca).}}
\maketitle
\begin{abstract}
Non-terrestrial networks have been attracting much interest from the industry and academia. Satellites and high altitude platform station (HAPS) systems are expected to be the key enablers of next-generation wireless networks.
In this paper, we introduce a novel downlink satellite communication (SatCom) model where free-space optical (FSO) communication is adopted between a satellite and a HAPS node. A hybrid FSO/radio-frequency (RF) transmission model is used between the HAPS node and ground station (GS). In the first phase of transmission, the satellite selects the HAPS node that provides the highest signal-to-noise ratio (SNR). In the second phase, the selected HAPS decodes and forwards the signal to the GS. To evaluate the performance of the proposed system, outage probability expressions are derived for exponentiated Weibull (EW) and shadowed-Rician fading models while considering the atmospheric turbulence, stratospheric attenuation, and attenuation due to scattering, path loss, and pointing errors. Additionally, asymptotic analysis is carried out and diversity gain is provided.
Furthermore, the impact of aperture averaging {technique}, temperature, and wind speed are investigated. We also provide some important guidelines that can be helpful for the design of practical HAPS-aided SatCom. Finally, the results show that the use of HAPS improves the system performance and that the proposed model performs better than all other existing models.
\end{abstract}
\section{Introduction}
{Low Earth} orbit (LEO) satellites have promising characteristics including high data rates, wide coverage, distance-independent services, and seamless connectivity for unserved and underserved communities. However, direct links between LEO satellites and ground stations (GSs) are highly susceptible to fading, atmospheric turbulence, path loss due to long transmitter-receiver distances, cloud formations, weather effects \cite{alam2020}, and masking effects caused by obstacles and shadowing \cite{guo2020p}. Thus, a cooperative model{,} in which a high altitude platform station (HAPS) aids satellite communication (SatCom) by using decode-and-forward (DF) and amplify-and-forward (AF) relaying strategies{,} can be a promising solution. According to the definition of the International Telecommunication Union (ITU), a HAPS system is a fixed object at an altitude of $20$ to $50$ km \cite{ITU2016}. However, the majority of recent deployments focused on an altitude of $18$ to $20$ km \cite{alam2020}. In SatCom, a HAPS system can function as an intermediate station to increase coverage and signal-to-noise ratio (SNR) while providing lower latency \cite{alam2020}. Furthermore, due to their small footprint, HAPS systems are suitable for low-latency applications and can deliver wireless communication services directly to terrestrial users \cite{9380673}. 

{The current architecture of SatCom relies on the microwave radio frequency (RF) band for most of the applications \cite{9530144}. However, RF communication suffers from limited capacity, congested spectrum, low bandwidth, and regulatory restrictions. Also, RF links are vulnerable to interception or jamming, which arises security problems. In this context, for} HAPS-aided SatCom, free-space optical (FSO) communication can be a potent enabler as it provides numerous advantages, including extremely high data rates, and it offers line-of-sight (LOS) connectivity over an unlicensed spectrum \cite{2020review}. Despite their unique features, FSO links are prone to significant variations in both phase and intensity of the received signal due to fluctuations in the index of refraction caused by variations in temperature and pressure mostly for longer communication distances \cite{andrews2005}. Furthermore, FSO communication is weather-dependent and highly affected by turbulence-induced fading and attenuation. More precisely, FSO links may be significantly degraded by fog {or} snow, whereas attenuation due to rain may be negligible \cite{weather}. {In addition, optical SatCom suffers from beam scintillation and beam wander effects, which are mainly caused by large-scale inhomogeneities in the atmosphere. Beam wander can be a significant factor in the uplink communication, as the beam size is much smaller than the turbulent eddies, which leads to beam displacement and link failure \cite{kaushal2016optical}. However, it can be negligible for downlink communication. Also, acquisition and pointing are challenging in FSO SatCom, which are mainly caused by devices vibration, platform jitter, or any type of stress in electronic or mechanical equipment. Thus, in order to avoid link failure, a LOS connection should be maintained between the transmitter and receiver \cite{kaushal2016optical}.}

Performance analyses of FSO communication have been reported in the literature using Log-normal, Gamma-Gamma, double Gamma-Gamma, and Mal\'{a}ga fading channels. Empirical studies, for their part, have proven that the exponentiated Weibull (EW) fading can be the best fit for different aperture sizes in all weather conditions, especially when aperture averaging is used, where the scintillation is spatially averaged over the aperture to mitigate the effects of atmospheric turbulence and improve the overall performance \cite{barrios2012}, \cite{erdogan2020}. Like aperture averaging, spatial diversity can be established between satellites and HAPS systems to provide reliable communication and mitigate the impact of atmospheric turbulence \cite{erdogan2020}.

Another way of reducing the effects of atmospheric turbulence is to incorporate the RF link in parallel with the FSO link to benefit from the complementary characteristics. Such hybrid RF/FSO communication can reap all the benefits of RF and FSO communication while minimizing adverse weather-dependent effects. In the literature, two methods of hybrid RF/FSO communication have been reported{;} soft switching and hard switching \cite{kazemi}. In hard switching, which is more practical, only one link might be active at a time. In this method, the FSO link is active initially, and the RF link acts as a backup {when the FSO becomes} unavailable. In case of soft switching, there is a simultaneous transmission on both links depending on their availability. Hybrid RF/FSO communication has been studied from different perspectives and in various scenarios in \cite{krishnan}, \cite{bag2018}, \cite{bag2019}. Furthermore, the authors in \cite{amirabadi} and \cite{2018selection} proposed a parallel transmission of the same information through RF and FSO links, while considering different diversity-combining techniques. {In the literature, different diversity combining techniques including selection combining (SC) and maximal ratio combining (MRC), are used to mitigate the impact of atmospheric turbulence on SatCom-based FSO communication \cite{ma2015performance}.}

The use of HAPS systems in hybrid RF/FSO communication has also been studied in the literature. In \cite{swaminathan}, the authors proposed a hybrid scheme for downlink SatCom using a HAPS node as an intermediate terminal between an LEO satellite and GS. In so doing, they derived the average symbol error probability (ASEP) while considering DF relaying and a backup RF channel in the second hop. More recently, in \cite{swamina}, the authors compared a single-hop hybrid RF/FSO SatCom with a dual-hop hybrid RF/FSO communication by using a HAPS node for the uplink. Their results showed that the hybrid RF/FSO outperforms the FSO systems and that the use of the HAPS node improves the performance of uplink SatCom.
{
Therefore, the main motivations of this paper are summarized as follows:
\begin{itemize}
   \item In the current literature, hybrid RF/FSO communication has been well investigated. However, these studies are mostly limited to horizontal terrestrial transmission, in which the distance is about 1-3 km. In addition, the performance analysis for downlink SatCom systems has been extensively carried out for single-hop FSO or RF communication and dual-hop mixed RF/FSO communication. Therefore, there is a significant gap in hybrid RF/FSO communication for satellite networks.
    \item In recent studies, a growing interest is witnessed in the use of HAPS node as a relay station in satellite-to-ground transmission to improve the SatCom performance.
    \item Recently, the same authors of \cite{swaminathan}, \cite{swamina}, \cite{9446153} are the first to investigate hybrid RF/FSO for HAPS-aided SatCom systems. However, their performance analyses mainly focus on symbol error probability (SEP) while assuming Gamma-Gamma distribution for FSO communication. In our work, we consider EW fading, which is shown to provide a better fit for larger apertures better than all other distributions \cite{barrios2013}. Furthermore, the impact of the aperture averaging technique, stratospheric attenuation, temperature variations, zero-boresight pointing errors, and rain attenuation have not been considered in their studies.
    \item To the best of the authors' knowledge, the HAPS selection for downlink SatCom has not yet been studied in the literature, and its potential to improve the system performance is still unknown.
\end{itemize}}
\begin{figure}[!t]
  \centering
    \includegraphics[width=3.7in, height=9cm]{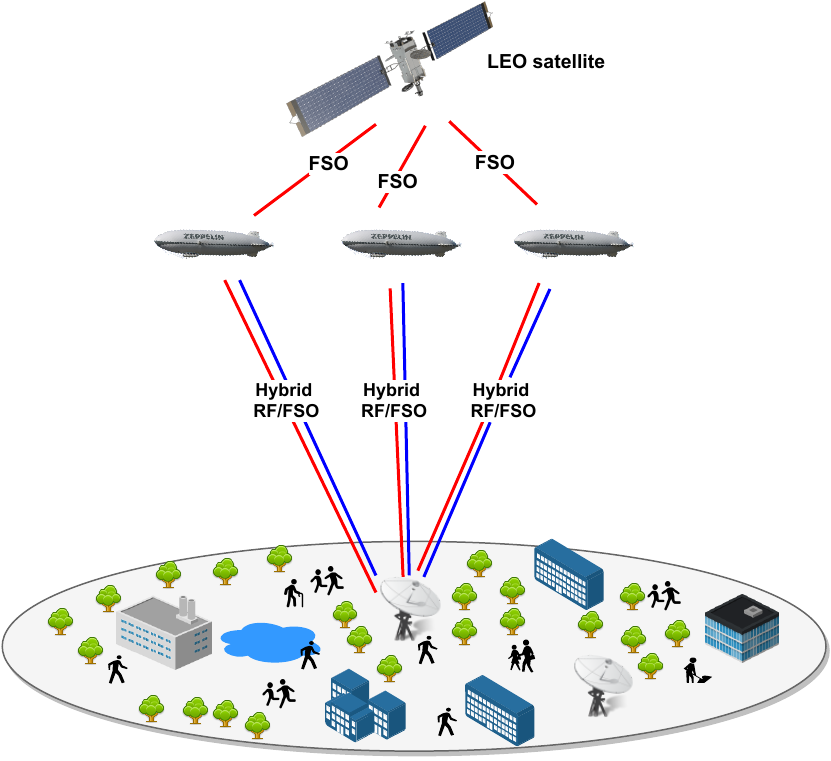}
    \caption{Illustration of the HAPS node selection for hybrid RF/FSO SatCom.}
  \label{fig:model}
\end{figure}
Our work in this paper differs from these studies by considering a HAPS-assisted SatCom model, where the HAPS with the best channel characteristics transmits the information to the GS by using hybrid RF/FSO communication. More precisely, the major contributions of this paper are summarized as follows: 
\begin{itemize}
\item We propose a novel model for downlink optical SatCom where the best HAPS node with the best channel characteristics is selected.

 \item To guarantee reliable communication, we consider FSO communication between the satellite and HAPS systems, and hybrid RF/FSO communication is adopted between the HAPS systems and GS.
 
 \item We introduce a novel stratospheric attenuation model for satellite-HAPS communication. Furthermore, for different weather conditions, we consider the effects of atmospheric turbulence, attenuation resulting from scattering, path loss, and pointing errors to provide realistic modeling for the proposed setup.

\item We investigate the effects of temperature and consider an aperture averaging technique to reduce fluctuations due to turbulence-induced fading and to mitigate the performance degrading effect of pointing errors.

\item We derive the outage probability expressions and validate them with Monte Carlo (MC) simulations. {Furthermore, asymptotic analysis is carried out to understand the behavior of the proposed system at high SNR.} We also provide some important design guidelines that can be helpful in the design of downlink SatCom systems.
\end{itemize}
This paper is organized as follows: The signals and system model are outlined in Section II. {Channel modeling and impairments are described in Section $\RN{3}$.} The mathematical expressions of outage probability {and asymptotic analysis are derived in Section $\RN{4}$}. In Section $\RN{5}$, numerical results are presented and discussed, followed by design guidelines. Finally, the conclusion is provided in Section $\RN{6}$.

\section{Signals and System Model}
{We consider a hybrid cooperative SatCom model consisting of an LEO satellite ($S$), a ground station ($G$), and $N$ DF  HAPS ($H$) nodes distributed randomly as depicted in Figure~1. The direct link between $S$ and $G$ is unavailable due to atmospheric attenuation or heavy shadowing. In this setup, $S$ selects the HAPS node denoted by ($H_J$) that can provide the best channel characteristics, based on the channel state information (CSI) feedback from the HAPS nodes. In the first hop, $S$ transmits its information to $H_J$ by using FSO communication. In the second hop, $H_J$ decodes the optical received signal and forwards it to $G$ by using hybrid RF/FSO communication. At the destination node $G$, the received signal with the highest SNR is selected to maximize the utilization of the channel spectrum. In the $S$ to $H_J$ communication, the Doppler shift effect can be reduced to enable reliable communication. Furthermore, due to the appealing quasi-stationary position of the HAPS node, tracking and precision problems can be ignored, as well as the introduction of the Doppler shift at the GS.} In this setup, the FSO links follow exponentiated Weibull (EW) fading, whereas the RF link is modeled with the shadowed-Rician distribution. Table $\RN{1}$ summarizes all the parameters used in the paper.
\begin{table}[t]
\caption{List of Parameters}
\label{notations}
\begin{center}
\begin{tabular}{ |l|l| } 
\hline
Parameter & Definition \\
\hline \hline
$N$ & Number of HAPS systems \\ 
$L$ & Propagation distance  \\ 
$\xi$ & Zenith angle   \\ 
$\mathcal{D}$ & Hard receiver aperture diameter   \\
$\Theta$ & Elevation angle \\
$u$ & RMS wind speed \\
$h_E$ & Height of the GS above mean sea level  \\
$h$ & Altitude  \\
$h_S$ & Altitude of the satellite \\
$h_H$ & Altitude of the HAPS node  \\
$\lambda$ & Wavelength \\
$\alpha, \beta$ & Shape parameters of the EW fading   \\
$\eta$  & Fading severity parameter of the EW fading   \\
$K$   &  Optical wave number  \\
$V$ &  Visibility  \\ 
$\mathcal{N}$ & Cloud number concentration  \\
$\omega$ & Geometrical attenuation coefficient \\
$\Psi$ & Stratospheric attenuation coefficient \\
$\mathcal{L_W}$ & Liquid water content \\
$\sigma_R^2$  & Rytov variance \\
$C_n^2$ & Refractive index constant \\
$\sigma_I^2$  & Scintillation index  \\
$\sigma_s$  & Jitter standard deviation  \\
$\mathcal{F}$  & Path loss  \\
$m$ & Nakagami-m fading severity parameter  \\
$\theta$ & Transmit divergence angle  \\
$\varphi_{rain}$ & Rain attenuation coefficient \\
$\varphi_{oxy}$ & Oxygen attenuation coefficient \\
$\mathcal{R}$ & Rain rate   \\
$n_f$ & Noise figure  \\
$T$ & Temperature  \\
$B$ & Bandwidth  \\
$\gamma_{th}$ &  Predefined threshold for acceptable communication quality  \\
\hline
\end{tabular}
\end{center}
\label{Tab2}
\end{table}
\normalsize
{\subsection{Satellite-HAPS Communication}}
Considering the communication between $S$ and $H_j$, stratospheric turbulence-induced fading can be caused by non-static stratospheric winds and temperature variations due to altitude and pressure. By considering stratospheric attenuation and stratospheric turbulence-induced fading, the received signal\footnote{{As widely adopted in the literature, we assume a time-invariant statistical model, so that we can obtain the outage probability in closed-form.}} {by} $H_j$ can be expressed as follows:
\begin{align}
y_{SH_j} =  \zeta \sqrt{P_S} I_{SH_j} x_S+ n_{H},
\label{EQN:2}
\end{align}
where $0\leq \zeta\leq1$ is the optical-to-electrical conversion coefficient, $P_S$ is the transmit power of $S$, $I_{SH_j}$ represents the irradiance of the FSO channel, which can be expressed as {$I_{SH_j}=  {I}_{SH_j}^a {I}_{SH_j}^t$, where  ${I}_{SH_j}^a$ stands for the stratospheric attenuation, ${I}_{SH_j}^t$ indicates the stratospheric turbulence,}
${x_S}$ denotes the transmitted signal, and $n_{H}$ is the zero-mean additive white Gaussian noise (AWGN) with one-sided power spectral density $N_{0}$. With respect to (\ref{EQN:2}), 
the instantaneous received SNR at $H_j$ can be expressed as
\begin{align}
\gamma_{SH_j}=\frac{ \zeta  {P_S} I_{SH_j}^2}{N_{0}}=\overline{\gamma}_{SH_j} I_{SH_j}^2,
\label{EQN:3}
\end{align}
where $\overline{\gamma}_{SH_j}=\frac{ \zeta  {P_S} }{N_{0}}$ is the average SNR with $\mathbb{E}[I_{SH_j}^2]=~1$.

In this setup, the HAPS selection is based on the satellite-HAPS channel quality, where the HAPS node with the best channel quality among $N$ HAPS is selected to maximize the instantaneous SNR between $S$ and $H_j$ as follows:
\vspace{-0.13cm}
\begin{align}
J = \arg \max_{j = 1,\ldots,N} [\gamma_{SH_j}], 
\label{EQN:4}
\end{align}
where $1\leq j\leq N$ shows the HAPS index. Considering the EW fading, the cumulative distribution function (CDF) of $\gamma_{SH_J}$ can be given as follows \cite{barrios2012}:

\vspace{-0.3cm}
\small
\begin{align}
 \begin{array}{c}
 F_{\gamma_{SH_J}}(\gamma)= {\displaystyle \prod_{j=1}^{N}} \Bigg( 1-\exp\left[ -\bigg( \frac{\gamma}{(\eta_{SH_j}{I}_{SH_j}^a)^2 \overline{\gamma}_{SH_j}}\right) ^{\frac{\beta_{SH_j}}{2}}\bigg]\Bigg)^{\alpha_{SH_j}},
 \label{EQN:6}
 \end{array}
\end{align}
\normalsize
where $\eta_{SH_j}$ is the scale parameter, $\alpha_{SH_j}$, and $\beta_{SH_j}$ present the shape parameters which are directly related to the atmosphere and the scintillation index. These parameters can be expressed as follows \cite{barrios2013}:
\begin{align}
\alpha_{SH_j}&=\frac{7.220 \times\sigma_{I_{SH_j}}^{2/3}}{\Gamma\left( 2.487\sigma_{I_{SH_j}}^{2/6} -0.104\right) }, \nonumber\\
\beta_{SH_j}&=1.012\left( \alpha_{SH_j}\sigma_{I_{SH_j}}^2\right) ^{-13/25}+0.142, \nonumber\\
\eta_{SH_j}&=\frac{1}{\alpha_{SH_j}\Gamma\left( 1+1/\beta_{SH_j}\right) \text{g}_{1}(\alpha_{SH_j},\beta_{SH_j})},  \nonumber\\
\label{EQN:7}
\end{align}
where $\Gamma(\cdot)$ indicates the Gamma function and $\text{g}_{1}(\alpha_{SH_j},\beta_{SH_j})$ is $\alpha_{SH_j}$ and $\beta_{SH_j}$ dependent constant variable given by \cite{erdogan2020}
\begin{align}
\text{g}_{1}(\alpha_{SH_j},\beta_{SH_j})=\sum_{k=0}^\infty \frac{(-1)^k \Gamma(\alpha_{SH_j})}{k! (k+1)^{1+1/\beta_{SH_j}}\Gamma(\alpha_{SH_j}-k)}.
\label{EQN:8}
\end{align}
$\sigma_{I_{SH_j}}^2$ is the scintillation index, which can be written as follows \cite[Sect. (12)]{andrews2005}:
\small
\begin{align}
\sigma_{I_{SH_j}}^2= \exp \left[ \frac{0.49 \sigma_{R_{SH_j}}^2}{(1+1.11 \sigma_{R_{SH_j}}^{12/5})^{7/6}} + \frac{0.51 \sigma_{R_{SH_j}}^2}{(1+0.69 \sigma_{R_{SH_j}}^{12/5})^{5/6}} \right] -1,
\label{EQN:9}
\end{align}
\normalsize
where $\sigma_{R_{SH_j}}^2$ denotes the Rytov variance given as follows \cite[Sect. (12)]{andrews2005}:
\small
\begin{align}
     \sigma_{R_{SH_j}}^2=2.25 K^{7/6} \sec^{11/6}(\xi_{SH_j}) \int_{h_H}^{h_S} C_{n_{SH_j}}^2(h)(h-h_H)^{5/6} dh.
     \label{EQN:10}
\end{align}
\normalsize
Here, $K=\frac{2\pi}{\lambda_{FSO}}$ denotes the optical wave number, $\xi_{SH_j}$ is the zenith angle, $h_S$ stands for the altitude of the satellite, $h_H$ indicates the height of the selected HAPS above ground level, and $C_{n_{SH_j}}^2(h)$ is the refractive-index parameter depending on the altitude $h$ expressed as follows \cite{ITUR}:
\begin{align}
     & C_{n_{SH_j}}^2(h)=8.148 \times 10^{-56} u_{SH_j}^2 h^{10} \exp(-h/1000)\nonumber\\
     & +2.7 \times10^{-16} \exp(-h/1500) +C_0 \exp(-h/100) ,
      \label{EQN:11}
\end{align}
where $u_{SH_j} =\sqrt{v_{{SH_j}}^2 +30.69 v_{{SH_j}} +348.91} $ represents the root-mean-square (RMS) of the wind speed, $v_{SH_j}$ is the wind speed in m/s at $H_j$, and $C_0$ is the nominal value of $C_{n_{SH_j}}^2(h)$ at the HAPS node in m$^{-2/3}$.
{\subsection{HAPS-Ground Station Communication}}
{\subsubsection{FSO Communication}}
Temperature and pressure gradients can cause variations in the atmosphere’s refractive index in the form of eddies causing atmospheric turbulence-induced fading. In the presence of atmospheric turbulence, the instantaneous SNR at $G$ can be expressed as follows:
\begin{align}
\gamma_{H_JG}^{FSO}=\frac{ \zeta P_{H_J} I_{H_JG}^2}{N_{0}}=\overline{\gamma}_{H_JG}^{FSO}
I_{H_JG}^2,
\end{align}
where $P_{H_J}$ represents the transmit power of the selected HAPS, $I_{H_JG}$ denotes the FSO channel gain (irradiance), which is given as $I_{H_JG}=I_{H_JG}^t I_{H_JG}^{a}$, where $I_{H_JG}^{t}$ indicates the atmospheric turbulence-induced fading {and $I_{H_JG}^{a}$ indicates the atmospheric attenuation}. Moreover, $\overline{\gamma}_{H_JG}^{FSO}=\frac{ \zeta  P_{H_J}}{N_{0}}$ is the average FSO SNR of the $H_J$ to $G$ link with  $\mathbb{E}[I_{H_JG}^2]=1$. Finally, the CDF of $\gamma_{H_JG}^{FSO}$ {can be expressed similarly to (\ref{EQN:6}), and the fading severity parameters can be obtained as in (\ref{EQN:7}) by replacing $SH_j$ subscript with $H_JG$ .}
{\subsubsection{RF Communication}}
Let $x_{H_J}$ denote the transmitted signal of $H_J$ with power $P_{H_J}$ through the RF link. The received signal at $G$ can be expressed as follows:
\begin{align}
y_{H_JG}=\sqrt{P_{H_J} \mathcal{F}_{H_JG}}f_{H_JG} x_{H_J} + n_{G},
\label{Eqsig}
\end{align}
where $n_{G}$ is the zero-mean AWGN with power spectral density $N_{0}$, $f_{H_JG}$ is the channel coefficient of the RF link that follows the shadowed-Rician fading, and $\mathcal{F}_{H_JG}$ is the path-loss model which can be expressed as \cite{kazemi}
\begin{align}
\mathcal{F}_{H_JG}[dB]&=G_{T} +G_R -20\log_{10} \left( \frac{4 \pi L_{H_JG}}{\lambda_{RF}}\right)\\ \nonumber
&-  \varphi_{oxy}L_{H_JG} -   \varphi _{rain}^{RF}L_{H_JG} ,
\end{align}
where $G_{T}$ and $G_R$ represent the gains of the transmitting and receiving antennas in dB, respectively. {$L_{H_JG}$ indicates the propagation distance between $H_J$ and $G$,} $\lambda_{RF}$ is the RF wavelength, $\varphi_{oxy}$ and $\varphi_{rain}^{RF}$ are the RF attenuation coefficients due to the oxygen and rain scattering \cite{touati}. In RF communication, the main attenuation factor is rain, where the corresponding attenuation increases linearly in relation to the rate of rainfall. Thus, the rain attenuation coefficient $\varphi_{rain}^{RF}$ (dB/km) can be expressed as \cite{ITUrain}
\begin{align}
\varphi_{rain}^{RF}= k_r \mathcal{R}^\varrho .
\end{align}
The parameters $k_r$ and $\varrho$ depend on the channel's wavelength $\lambda_{RF}$(GHz) and can be given as follows \cite{ITUrain}:
\begin{align}
k_r&=[k_H +k_V +(k_H-k_V)\cos^2\Theta \cos 2\iota]/2, \nonumber \\
\varrho&=[k_H\varrho_H+k_V\varrho_V +(k_H\varrho_H-k_V\varrho_V)\cos^2\Theta \cos 2\iota]/2k_r,
\end{align}
where the constants $k_H$, $k_V$, $\varrho_H$, and $\varrho_V$ are given in \cite{ITUrain} and $\iota$ is the polarization tilt angle \cite{ITUrain}. Therefore, with the help of (\ref{Eqsig}), the instantaneous SNR at $G$ can be written as follows:
\begin{align}
\gamma_{H_JG}^{RF}=\frac{P_{H_J} \mathcal{F}_{H_JG}|f_{H_JG}|^2}{N_{0}} = \overline{\gamma}_{H_JG}^{RF}|f_{H_JG}|^2.
\end{align}
where $ \overline{\gamma}_{H_JG}^{RF} = \frac{P_{H_J} \mathcal{F}_{H_JG}}{N_{0}}$ is the average SNR for the RF link between $H_J$ and $G$ with $\mathbb{E}[|f_{H_JG}|^2]=1$. Furthermore, the PDF of the received SNR for the RF link is given by \cite{2019physical}:
\begin{align}
f_{\gamma_{H_JG}}^{RF} (\gamma)&= \sum_{l=0}^{m -1} \frac{\mu(1-m)_l (-\delta)^l }{(\overline{\gamma}_{H_JG}^{RF})^{l+1} (l!)^2} ({\gamma})^l\exp(-\vartheta {\gamma}),
\end{align}
where $\mu=\frac{1}{2b}(\frac{2 b m}{2 b m + \Omega})^{m}$, $ \delta=\frac{\Omega}{2b(2b m + \Omega)}$, { $\vartheta=\frac{\nu - \delta}{\overline{\gamma}_{H_JG}^{RF}}$}, and $\nu=~\frac{1}{2b}$ with $m$ is a positive integer representing the Nakagami-m fading parameter of the corresponding link. Furthermore, $\Omega$ and $2b$ are the average power of the LOS component and multi-path component, and $(\cdot)_l$ indicates the Pochhammer symbol.
{
\section{Attenuation, Pointing Loss, and Temperature Variations}}
{\subsection{Satellite-HAPS Communication }}
\subsubsection{Stratospheric Attenuation}
In addition to its low costs and faster services, at the HAPS level, there are no clouds, which means clean solar energy without atmospheric pollution \cite{aragon2008}. However, for long-distance communication, the possibility of volcanic eruptions and resulting aerosol emissions needs to be considered, given that such aerosols can penetrate the stratosphere \cite{2010optical}. Moreover, stratospheric attenuation caused by molecular absorption and scattering by droplets can take place \cite{giggenbach}. Considering the stratospheric conditions, polar clouds can cause temperature differences between HAPS and satellite, and this can result in fluctuations in the optical beam. In optical communication, the stratospheric attenuation can be modeled with Beer-Lambert law as follows:
\begin{align}
    {I}_{SH_j}^a=\exp(-\Psi_{SH_j} L_{SH_j}),
    \label{EQN:1}
\end{align}
where $\Psi_{SH_j}$ represents the attenuation factor between the satellite and the HAPS systems, and $L_{SH_j}$ is the propagation distance between $S$ and $H_j$ \cite{kazemi}. In Table $\RN{2}$, we present stratospheric aerosol models for different levels of volcanic activity at {the} optical wavelength $\lambda_{FSO}=1550$ nm \cite{giggenbach}.
\begin{table}[t]
\caption{Stratospheric attenuation coefficient for different stratospheric aerosol models at $\lambda_{FSO}=1550$ nm for HAPS altitude 19 km. }
\label{notations}
\begin{center}
\begin{tabular}{ |l| l|  } 
\hline
Stratospheric aerosol model & Attenuation coefficient $\Psi$ (km$^{-1}$) \\
\hline
\hline
Extreme volcanic & $2\times 10^{-1}$ \\
High volcanic & $  5\times10^{-2}$ \\
Moderate volcanic  & $  8\times 10^{-3}$\\ 
Background volcanic & $  10^{-4}$\\
\hline
\end{tabular}
\end{center}
\label{Tab1}
\end{table}
\subsubsection{Pointing Errors}
For FSO communication, another critical impairment consists of beam-pointing errors, which significantly affect the performance of networks, especially over large distances. Due to vibrations in the transmitter telescope and thermal expansion, a misalignment between the transmitter and receiver can occur. Pointing errors come down to two issues. First, the boresight, which is the fixed displacement between the beam center and center of the detector. Second, jitter, which represents the random offset of the beam center at the detector plane \cite{2018performance}. In the presence of pointing errors, the irradiance of the FSO channel can be expressed as $I_{SH_j}=  {I}_{SH_j}^a {I}_{SH_j}^t {I}_{SH_j}^p$, where ${I}_{SH_j}^p$ indicates the pointing errors component.
In our model, we will assume zero-boresight pointing errors for the link between $S$ {and} all instances of $H$. Hence, the PDF of $I^p_{SH_j}$ can be given as follows \cite{yang2014free}:
\begin{align}
\label{error}
    f_{I^p_{SH_j}}(I^p)=\frac{g^2 \exp(\frac{-s^2}{2 \sigma_s^2})}{A_0^{g^2}} (I^p)^{g^2 -1} I_0 \Bigg( \frac{s}{\sigma_s^2} \sqrt{\frac{-w^2_{eq} \ln{\frac{I^p}{A_0}}}{2}} \Bigg),
\end{align}
where the parameter $g=w_{eq}/(2\sigma_{s})$ is the ratio between the equivalent beam $w_{eq}$ and the jitter standard deviation $\sigma_{s}$, where $w_{eq}^2=w_{z}^2 \sqrt{\pi} \text{erf}(y)/(2ye^{(-y^2)})$. $y=\sqrt{\pi/2\varpi}/w_{z}$ indicates the ratio of the aperture radius $\varpi$ and the beam-width $w_{z}$ at distance $z$, $w_{z}=\theta z$ with $\theta$ is the beam divergence angle, and $\text{erf}(\cdot)$ indicates the error function. Moreover, $s$ denotes the boresight, which is considered to be zero in our case. Finally, $A_{0}=[\text{erf}(y)]^2$ defines the gathered optical power for a zero difference between the optical spot center and the detector center, and $I_0(x)$ defines the modified Bessel function of the first kind with order zero \cite{yang2014free}. 
\subsubsection{Aperture Averaging}
For downlink optical communication, when the receiving aperture is lower than the correlation width of irradiance fluctuations, the turbulence-induced signal fluctuations can deteriorate the system performance. Hence, aperture averaging takes place, and increasing the aperture size not only improves the signal level but also reduces the fluctuations in the received signal. More specifically, the aperture size-dependent scintillation index can be given as follows \cite[Sect. (12)]{andrews2005}: 

\vspace{-0.3cm}
\small
\begin{align}
\label{aperture}
\sigma_{I_{SH_j}}^2&=8.7k^{7/6}(h_S-h_H)^{5/6} \sec^{11/6}(\xi_{SH_j}) \times\Re \Bigg\lbrace \int_{h_H}^{h_S}  C_{n_{SH_j}}^2 (h) \nonumber\\
&\times \left[ \left( \frac{k \mathcal{D}_{SH_j}^2}{16 L_{SH_j}} + i \frac{h-h_H}{h_S - h_H}\right)^{5/6} -\left( \frac{k \mathcal{D}_{SH_j}^2}{16 L_{SH_j}} \right)^{5/6}
\right] \Bigg\rbrace  dh,
\end{align}
\normalsize
where $\mathcal{D}_{SH_j}$ is the hard aperture diameter of the HAPS node in meter.
\subsection{HAPS-Ground Station Communication}
In $H_J$ to $G$ communication, we adopt hybrid RF/FSO communication, where RF or FSO communication can be selected at ground level depending on the channel characteristics. In other words, $G$ chooses the best link that maximizes the instantaneous SNR between $H_J$ and $G$.
\subsubsection{Atmospheric Attenuation}
The main problem in the optical wireless links is attenuation resulting from scattering and absorption. The scattering of optical signals is mainly caused by weather conditions such as clouds, fog, snow, and rain \cite{weather}.

In optical communication, Mie scattering is considered to be one of the main sources of signal loss in downlink channels operating at frequencies below $375$ THz. It affects the signal when the wavelength is equal to the diameter of the particles in the medium. The following formula, which is used to show the effect of Mie scattering, is suitable for ground stations located at altitudes of $0<h_E< 5$ km above the mean sea level. First, we calculate the wavelength-dependent empirical coefficients as follows \cite{ITUR2003}:
\begin{align}
a&= -0.000545\lambda_{FSO}^2 +0.002\lambda_{FSO}-0.0038 \nonumber \\
b&=0.00628\lambda_{FSO}^2-0.0232 \lambda_{FSO}+0.0439 \nonumber\\
c&=-0.028\lambda_{FSO}^2 +0.101\lambda_{FSO}-0.18 \nonumber \\
d&=-0.228\lambda_{FSO}^3+0.922\lambda_{FSO}^2-1.26\lambda_{FSO}+0.719,
\end{align}
where $h_E$ indicates the altitude of $G$ above sea level.
Then, the extinction ratio can be expressed as follows \cite{ITUR2003}:
\begin{align}
\tau=ah_E^3 +bh_E^2+ch_E+d,
\end{align}
and the atmospheric attenuation due to Mie scattering can be given as follows:
\begin{align}
I^{m}_{H_JG}=\exp\left( -\frac{\tau}{\sin(\Theta)} \right) ,
\end{align}
where $\Theta$ is the elevation angle of the GS.

In optical communication, geometrical scattering can also deteriorate the signal in the atmosphere. In geometrical scattering, fog and cloud-induced fading are the primary causes of FSO communication deterioration. To estimate the attenuation based on the visibility range parameters, the well-known Kim's model can be used to define the attenuation coefficient as \cite{weather}
\begin{align}
    \omega=\frac{3.91}{V}\left( {\frac{\lambda_{FSO}}{550}}\right) ^{-x},
          \end{align}
where $V$ defines the visibility range in km, and $x$ implies the particle size coefficient of scattering given by the Kim model as follows:
  \begin{align}
         x= \begin{cases}   
           1.6  & \hspace{0.2cm} V > 50 \\
          1.3  &  \hspace{0.2cm} 6 < V <50\\
          0.16V+0.34  & \hspace{0.2cm} 1 < V < 6   \\
          V-0.5  & \hspace{0.2cm}  0.5 < V < 1  \\
          0   &  \hspace{0.2cm}   V < 0.5  .
\end{cases}
\end{align}
   \begin{table}[!t]
   \renewcommand{\arraystretch}{1.1}
   \centering
\caption{Atmospheric attenuation and visibility parameters for different attenuation and visibility parameters for different fog conditions at $\lambda_{FSO}=1550  \text{nm}$.} 
\label{tab1}
\begin{tabular}{|c|c|c|}
\hline  Fog & $V$ (km) & Attenuation coefficient $\omega$ (dB/km) \\
\hline  Dense & $0.05$ & $339.62$ \\
\hline  Thick & $0.20$ & $84.90$\\
\hline  Moderate & $0.50$ & $33.96$ \\
\hline  Light & $0.77$ & $16.67$ \\
\hline  Thin & $1.90$ & $4.59$ \\
\hline
\end{tabular}
\end{table}
   \begin{table}[!t]
   \renewcommand{\arraystretch}{1.1}
   \centering
\caption{Geometrical scattering parameters for various types of clouds at $\lambda_{FSO}=1550\text{nm}$.} 
\label{tab1}
\begin{tabular}{|c|c|c|c|}
\hline  Cloud type & $\mathcal{N}$ ({cm}$^{-3}$) & $\mathcal{LW}$ (g/m$^{-3}$) & $V$ (km) \\
\hline  Cumulus & 250 & 1.0 & 0.0280 \\
\hline  Stratus & 250 & 0.29 & 0.0626 \\
\hline  Stratocumulus & 250 & 0.15 & 0.0959 \\
\hline  Altostratus & 400 & 0.41 & 0.0369 \\
\hline  Nimbostratus & 200 & 0.65 & 0.0429 \\
\hline  Cirrus & 0.025 & 0.06405 & 64.66 \\
\hline  Thin cirrus & 0.5 & 3.128 $\times$ 10$^{\text{-4}}$& 290.69\\
\hline
\end{tabular}
\end{table}
In Table $\RN{3}$, the visibility and attenuation coefficient parameters are presented for different fog conditions. Based on this model, for different cloud types, the visibility can be given by using the liquid water content ($\mathcal{L_W}$) and cloud number concentration ($\mathcal{N}$) as follows \cite{awan2009}:
 \begin{align}
 V=\frac{1.002}{(\mathcal{L_W}\mathcal{N})^{0.6473}}.
    \end{align}
The corresponding parameters are summarized in Table $\RN{4}$. Accordingly, the geometrical attenuation can be given by using the Beer-Lambert law as $I^{g}_{H_JG}=\exp(-\omega L_{H_JG})$. Hence, the total atmospheric attenuation at ground level can be expressed {as} \cite{johari2017}:
\begin{align}
   I_{H_JG}^{a}=I_{H_JG}^{m}I_{H_JG}^{g}.
\end{align}
Among the different atmospheric effects on the FSO link, rain is the weakest attenuation factor. However, the size of rain droplets increases when the rainfall rate increases, so it may cause refraction and reflection. Considering the FSO communication, the specific rain attenuation coefficient can be expressed on the basis of the rainfall rate $\mathcal{R}$ (mm/h) as follows \cite{weather}:
\begin{align}
      \varphi_{rain}^{FSO}=1.076 \mathcal{R}^{0.67}. 
\end{align}
Therefore, the rain attenuation can be obtained using the Beer-Lambert law as $I_{H_JG}^{rain}=\exp(-  \varphi_{rain}^{FSO} L_{H_JG})$. Thus, in the presence of rain, the total attenuation can be considered as $ I_{H_JG}^{a}=I_{H_JG}^{m}I_{H_JG}^{g} I_{H_JG}^{rain}$.
{
\subsubsection{Aperture Averaging}
Aperture averaging technique is also considered in the second-hop link to improve the communication from ${H_J}$ to $G$. Therefore, the scintillation index dependent aperture diameter can be similarly expressed as in (\ref{aperture}) by just changing the subscripts as $\sigma_{I_{H_JG}}^2$ for the scintillation index and $\mathcal{D}_{H_JG}$ for the hard aperture.}
{
\subsubsection{Pointing Errors}
In this subsection, pointing errors due to misalignment between $H_J$ and $G$ is taken into consideration. Thus, in the presence of zero-boresight pointing errors for ${H_J}$ to $G$ communication, the irradiance of the channel can be written as $I_{H_JG}=I_{H_JG}^{t}I_{H_JG}^{a}I_{H_JG}^{p}$, with $I_{H_JG}^{p}$ is the pointing errors component. The PDF of $I_{H_JG}^{p}$ can be written similarly as in (\ref{error}).}

\subsection{Impact of the Temperature Variations}
The Earth's atmosphere extends up to $700$ km above ground level and is divided into four distinct layers on the basis of temperature. SatCom can be affected by the thermal noise, which varies with altitude. For our proposed model, we consider the troposphere and stratosphere layers \cite[Sect. (1)]{andrews2005}.
\begin{itemize}
\item Troposphere: This layer extends up to 11 km and contains 75$\%$ of the Earth's atmospheric mass. The maximum air temperature takes place near the ground and decreases up to -55°C with an increase of altitude.
\item Stratosphere: This layer starts at 20 km and extends up to 48 km. The air temperature level decreases with an increase of the altitude starting from -55°C.
\end{itemize}
We analyze the impact of the thermal noise associated with these layers. We show that the use of the HAPS node improves the system's performance as the link between the satellite and the HAPS node is less affected by the noise. The noise power $N_0$ can be given as $N_0=P_n n_f$, where $n_f$ is the noise figure of the receiver and $P_n$ is given as follows \cite{swamina}:
\begin{align}
P_n(dB)= k+ T+ B,
\label{Eq:Noise}
\end{align}
where $k=-228.6$ dBW/K/Hz represents the Boltzmann's constant, $T$ is the system noise temperature in dBK, and $B$ denotes the noise bandwidth in dBHz. Please note that (\ref{Eq:Noise}) can be used either for a HAPS node or GS, depending on the temperature.
{
\section{Performance Analysis}
\subsection{Outage Probability}
}
The outage probability of a communication channel is defined as the probability of the instantaneous SNR falling below a predefined threshold $\gamma_{th}$. This can be expressed as follows \cite{2018outage}:
\begin{align}
\label{outage}
    P_{out}&=\Pr[\gamma_0 \leq \gamma_{th}]\\ \nonumber
&=F_{\gamma_0}(\gamma_{th}),
\end{align}
where $F_{\gamma_0}(\gamma_{th})$ is the CDF of the end-to-end SNR at $G$, which can be given as:
\begin{align}
    \gamma_0=\min(\gamma_{SH_J}, \gamma_{H_JG}),
\end{align}
where $\gamma_{H_JG}=\max (\gamma_{H_JG}^{FSO},\gamma_{H_JG}^{RF})$ is the output SNR of the SC at $G$, and the CDF of the $\gamma_0$ can be written as follows:
 \begin{align}
 \label{end-to-endSNR}
 &F_{\gamma_0} \left( \gamma \right) = 1- \Pr [\gamma_{SH_J} > \gamma]  \Pr [\gamma_{H_JG}> \gamma] \nonumber\\ 
 & =1-\left(1- F_{\gamma_{SH_J}} \left( \gamma \right) \right) \left(1- F_{\gamma_{H_JG}} \left( \gamma \right) \right) \nonumber\\ 
 & =1-\left(1- F_{\gamma_{SH_J}} \left( \gamma \right) \right) \left(1- \left(F_{\gamma_{H_JG}^{FSO}}(\gamma) F_{\gamma_{H_JG}^{RF}}(\gamma) \right) \right),
  \end{align}  
where $F_{\gamma_{H_JG}^{FSO}}(\gamma)$ and $F_{\gamma_{H_JG}^{RF}}(\gamma)$ are the CDF of $\gamma_{H_JG}^{FSO}$ and $\gamma_{H_JG}^{RF}$, respectively, given as follows:
\small
 \begin{align}
 \label{CDF_FSO}
 F_{\gamma_{H_JG}^{FSO}}(\gamma)&=\sum_{\rho=0}^{\infty} \left( \begin{array}{c} \alpha_{H_JG} \\
 \rho
   \end{array}  \right) 
   (-1)^{\rho} \\ \nonumber
  &\times \exp\left[ -\rho \left( \frac{\gamma}{(\eta_{H_JG} I_{H_JG}^{a})^2 \overline{\gamma}_{H_JG}^{FSO}} \right) ^{\frac{\beta_{H_JG}}{2}}\right] ,     
  \end{align}
  
  \begin{align}
  \label{CDF_RF}
F_{\gamma_{H_JG}^{RF}} (\gamma)&= 1- \sum_{l=0}^{m -1} \sum_{q=0}^{l} \frac{\mu(1-m)_l(-\delta)^l }{q! \vartheta^{l-q+1}(\overline{\gamma}_{H_JG}^{RF})^{l+1} l!} \\ \nonumber
&\times (\gamma)^q\exp(-\vartheta \gamma).
\end{align}
\normalsize
Furthermore, $F_{\gamma_{SH_J}}(\gamma)$ is the CDF of $\gamma_{SH_J}$, which can be expressed as given in (\ref{EQN:6}) in the absence of pointing errors. However, in the presence of pointing errors, $F_{\gamma_{SH_J}}(\gamma)$ can be {obtained as} \cite{2018performance}:
\begin{align}
\label{CDFPE}
F_{\gamma_{SH_J}}(\gamma)&= {\displaystyle \prod_{j=1}^{N}} \Bigg(\frac{\alpha_{SH_j} g^2}{\beta_{SH_j}} \left( \frac{1}{\eta_{SH_j} A_{0}} \sqrt{\frac{\gamma}{\overline{\gamma}_{SH_j} ({I}_{SH_j}^a)^2}}\right) ^{g^2} \nonumber\\ 
&\times \sum_{i=0}^{\infty} T_{2}(i) G_{2,3}^{2,1} \left( T_{3}(i) \middle \vert \begin{array}{c}
1-T_{1}  , 1 \\
0,1-T_{1} , -T_{1} 
\end{array} 
\right) \Bigg),
\end{align}
\normalsize
where $G_{p,q}^{m,n} \Big( x\hspace{0.1cm} \Big| \begin{matrix} a_1,...,a_p\\  b_1,...,b_q \end{matrix} \Big)$ denotes the Meijer G-function \cite[eqn. 07.34.02.0001.01]{Wolform},  $T_{1}=~g^2/\beta_{SH_j}$, $T_{2}(i)=(-1)^i \Gamma(\alpha_{SH_j}) /[i!\Gamma(\alpha_{SH_j} -i)
(1+i)^{1-T_{1}}]$, and
$T_{3}(i)=(1+i)\left( \frac{1}{\eta_{SH_j} A_{0}} \sqrt{\frac{\gamma}{\overline{\gamma}_{SH_j} ({I}_{SH_j}^a)^2}}\right) ^{\beta_{SH_j}}$. 

{Similarly, in the presence of zero-boresight pointing errors from $H_J$ to $G$, the CDF of $\gamma_{H_JG}^{FSO}$ can be written as in (\ref{CDFPE}) after changing the subscripts by $H_JG$.}

Finally, by substituting (\ref{CDF_FSO}) and (\ref{CDF_RF}) into (\ref{end-to-endSNR}), then into (\ref{outage}), the final expression of the outage probability can be obtained, as can be seen at the top of the next page.
\small
\begin{figure*}[t!]
\begin{align}
  &P_{out} (\gamma_{th})=  \prod_{j=1}^{N}  \sum_{\rho=0}^{\infty}  \binom{\alpha_{SH_j}}{\rho}
   (-1)^{\rho} \exp\left( -\rho \left( \frac{\gamma_{th}}{(\eta_{SH_j}I_{SH_j}^a)^2 (P_S /N_0)} \right) ^{\frac{\beta_{SH_j}}{2}}\right)  
   + \left[  \sum_{p=0}^{\infty} \binom{\alpha_{H_JG}}{p} (-1)^{p} \nonumber  \right.\\
  & \left.\times 
   \exp\left( -p \left( \frac{\gamma_{th}} {(\eta_{H_JG} I_{H_JG}^{a})^2 (P_{H_J}/N_0)} \right) ^{\frac{\beta_{H_JG}}{2}}\right) \times \Bigg(1- \sum_{l=0}^{m -1} \sum_{q=0}^{l} \frac{\mu(1-m)_l.(-\delta)^l }{q! \vartheta^{l-q+1}(P_{H_J} \mathcal{F}_{H_JG}/N_0)^{l+1} l!}  \gamma_{th}^q\exp(-\vartheta \gamma_{th}) \Bigg) \right] \nonumber\\
  & - \left[ \prod_{j=1}^{N} \sum_{\rho=0}^{\infty} \binom{\alpha_{SH_j}}{\rho}
   (-1)^{\rho}  \exp \left( -\rho \left( \frac{\gamma_{th}}{(\eta_{SH_j}I_{SH_j}^a)^2 (P_S /N_0)} \right) ^{\frac{\beta_{SH_j}}{2}} \right)
\times \Bigg( 1- \sum_{l=0}^{m -1} \sum_{q=0}^{l} \frac{\mu(1-m)_l.(-\delta)^l }{q! \vartheta^{l-q+1}(P_{H_J} \mathcal{F}_{H_JG}/N_0)^{l+1} l!} \nonumber \right. \\
& \left. \times \gamma_{th}^q
\exp(-\vartheta \gamma_{th}) \Bigg)
\times \sum_{p=0}^{\infty} \binom{\alpha_{H_JG}}{p} (-1)^{p}
   \exp\left( -p \left( \frac{\gamma_{th}} {(\eta_{H_JG} I_{H_JG}^{a})^2 (P_{H_J}/N_0)} \right) ^{\frac{\beta_{H_JG}}{2}} \right) 
   \right] .
\label{EQN:47}
\end{align}
\hrulefill
\end{figure*}
\normalsize

{
\subsection{High SNR Analysis}
In this subsection, the asymptotic expressions of outage probability are derived to get the diversity order of the proposed system. 
Similar to (\ref{end-to-endSNR}), the outage probability at higher SNR can be written as
\begin{align}
\label{Poutinf}
    P_{\text{out}}^{\infty}&=F_{\gamma_{SH_J}}^{\infty} (\gamma_{th})+F_{\gamma_{H_JG}^{FSO}}^{\infty}(\gamma_{th}) F_{\gamma_{H_JG}^{RF}}^{\infty} (\gamma_{th})\nonumber\\
    &- F_{\gamma_{SH_J}}^{\infty}(\gamma_{th}) F_{\gamma_{H_JG}^{FSO}}^{\infty}(\gamma_{th}) F_{\gamma_{H_JG}^{RF}}^{\infty}(\gamma_{th})\nonumber \\
    &\approx F_{\gamma_{SH_J}}^{\infty}(\gamma_{th})+F_{\gamma_{H_JG}^{FSO}}^{\infty}(\gamma_{th}) F_{\gamma_{H_JG}^{RF}}^{\infty}(\gamma_{th}).
\end{align}
The negative term in (\ref{Poutinf}) is neglected as its value is very small compared to the sum of the other terms. By using the Taylor series approximation of $\exp(-x/a) \simeq 1-x/a$, and after few manipulations, $F_{\gamma_{SH_J}}^{\infty}(\gamma_{th})$ can be written as 
\begin{align}
\label{asyCDF}
   F_{\gamma_{SH_J}}^{\infty}(\gamma_{th})= \prod_{j=1}^{N}\left[ \left( \frac{\gamma_{th}}{{(\eta_{SH_j}} I^a_{SH_j})^2 \overline{\gamma}_{SH_j}}\right) ^{\frac{\alpha_{SH_j}\beta_{SH_j}}{2}}\right] .
\end{align}
Note that in $H_J$ to $G$ FSO communication, $F_{\gamma_{H_JG}^{FSO}}^{\infty}(\gamma_{th})$ can be obtained similarly as in (\ref{asyCDF}) after changing the subscripts as \small
$F_{\gamma_{H_JG}^{FSO}}^{\infty}(\gamma_{th})= \left( \frac{\gamma_{th}} {(\eta_{H_JG} I^a_{H_JG})^2 \overline{\gamma}_{H_JG}^{FSO}}\right) ^{{\alpha_{H_JG}\beta_{H_JG}}/{2}}$\normalsize. 
On the contrary, for $H_J$ to $G$ RF communication, to obtain $F_{\gamma_{H_JG}^{RF}}^{\infty}(\gamma_{th})$, we apply Maclaurin series expansion \cite{2014table} for the exponential function and consider only the first term as the higher-order terms are negligible. Therefore, $F_{\gamma_{H_JG}^{RF}}^{\infty}(\gamma_{th})$ can be written as 
\begin{align}
 F_{\gamma_{H_JG}^{RF}}^{\infty} (\gamma_{th})\simeq \mu \gamma_{th}\frac{ 1}{\overline{\gamma}_{H_JG}^{RF}}.
 \end{align}
 Furthermore, at high SNR values, the outage probability can be written as $P_{\text{out}}^{\infty}=k (\overline{\gamma})^{-\mathcal{G}_d}$, where $k$ is a constant variable, which defines the coding gain of the system. The diversity order $\mathcal{G}_d$ defines the slope of the outage probability curve. In the case when the average SNRs of all links tend to infinity, the diversity gain is obtained as $\mathcal{G}_d=\min \Big( N \frac{\alpha_{SH_j}\beta_{SH_j}}{2}  , \max(\frac{\alpha_{H_JG}\beta_{H_JG}}{2}, 1) \Big)$.
 Finally, the asymptotic outage probability $P_{\text{out}}^{\infty}$ can be easily obtained.
}
\section{Numerical Results and Discussion}
In this section, we first validate the theoretical results with the MC simulations. Then, we evaluate the outage probability of our system model under different weather conditions. In the simulations, the effects of aperture averaging, pointing errors, wind speed, and different levels of thermal noise are investigated in terms of outage probability. Furthermore, we assume that all HAPS systems experience the same atmospheric conditions without losing the generality, and we assume equal transmit power at $S$ and $H_j$. The fading severity parameters for the RF link, which is modeled as a shadowed-Rician fading channel, are simulated depending on different shadowing severity levels{;} frequent heavy shadowing ($m$ = 1.0, $b$ = 0.063, $\Omega$ = 8.94 $\times$ 10$^{\text{-4}}$), average shadowing ($m$ = 10, $b$ = 0.126, $\Omega$ = 0.835), and infrequent light shadowing ($m$ = 19, $b$ = 0.158, $\Omega$=1.29) \cite{2019physical}. In addition, the following rain-rate parameters are set: light rain ($\mathcal{R} = $2.5 mm/h), moderate rain ($\mathcal{R}=$12.5 mm/h), and heavy rain ($\mathcal{R} = $25 mm/h) \cite{kazemi}. Moreover, for the FSO link between $H_J$ and $G$, the atmospheric turbulence parameters are set to ($\alpha_{H_JG}$ = 3.3419, $\beta_{H_JG}$ = 2.3131, $\eta_{H_JG}$ = 0.78693) for $u_{H_JG}$ = 21 m/s {in the absence of aperture averaging technique}, whereas the stratospheric turbulence parameters are set to {($\alpha_{SH_j}$ = 1.5825, $\beta_{SH_j}$ = 8.9870, $\eta_{SH_j}$ = 1.0025) for $u_{SH_j}$ = 65 m/s} without aperture averaging. In all simulations, we assume the sky to be homogeneous and the atmospheric attenuation to change in function of altitude. Finally, the outage probability is plotted relative to the transmit power at a threshold $\gamma_{th}=$7 dB. Table $\RN{5}$ provides the simulation parameters used in the numerical results section. 
 \begin{table}[!t]
   \renewcommand{\arraystretch}{1.1}
   \centering
\caption{Parameters of FSO and RF links} 
\label{tab1}
\begin{tabular}{|c|c|}
 \hline
 \multicolumn{2}{|c|}{\textbf{Satellite-HAPS (FSO)}} \\
\hline  Parameter & Value  \\
\hline Zenith angle ($\xi_{SH_j}$) & 65°  \\
\hline  Wind speed ($u_{SH_j}$) & 65 m/s\\
\hline Optical wavelength ($\lambda_{FSO}$) & 1550 nm \\
\hline Temperature ($T$) &  -55 °C \\
\hline Noise figure ($n_f$) &  1 dB \\
\hline Satellite height ($h_S$) &  500 km \\
\hline HAPS altitude ($h_H$) & 19 km  \\
\hline Stratospheric attenuation ($\Psi$) & 2.15$ \times$ 10$^{\text{-1}}$  \\
\hline Bandwidth (B) & $0.5$ GHz  \\
\hline Nominal value ($C_0$) & 10$^{\text{-18}}$  \\
\hline \hline 
 \multicolumn{2}{|c|}{\textbf{HAPS-GS (FSO)}} \\
\hline  Zenith angle ($\xi_{H_JG}$) & 20° \\
\hline  Wind speed ($u_{H_JG}$) & 21 m/s\\
\hline Optical wavelength ($\lambda_{FSO}$) & 1550 nm \\
\hline  Elevation above sea level ($h_E$) & 0.8 km \\
\hline Nominal value ($C_0$) & 1.7$ \times$ 10$^{\text{-14}}$  \\
\hline \hline
 \multicolumn{2}{|c|}{\textbf{ HAPS-GS (RF)}} \\
 \hline RF wavelength ($\lambda_{RF}$) & 40 GHz\\
\hline  Transmitter gain ($G_T$) & 45 dB\\
\hline  Receiver gain ($G_R$) & 45 dB\\
\hline  Polarization tilt angle ($\iota$) & 45°\\
\hline  Oxygen scattering ($\varphi_{oxy}$) & 0.1 dB/km \cite{el2011trends}\\
\hline \hline 
 \multicolumn{2}{|c|}{\textbf{ Common Parameters for HAPS-GS RF and FSO}} \\
\hline Bandwidth (B) & 0.5 GHz  \\
\hline Temperature ($T$) &  18 °C \\
\hline Noise figure ($n_f$) &  1 dB \\
\hline Threshold ($\gamma_{th}$) &  7 dB \\
\hline 
\end{tabular}
\end{table}
\subsection{Verification of the Theoretical Expressions}

\begin{figure}[!t]
  \centering
    \includegraphics[width=3.6in]{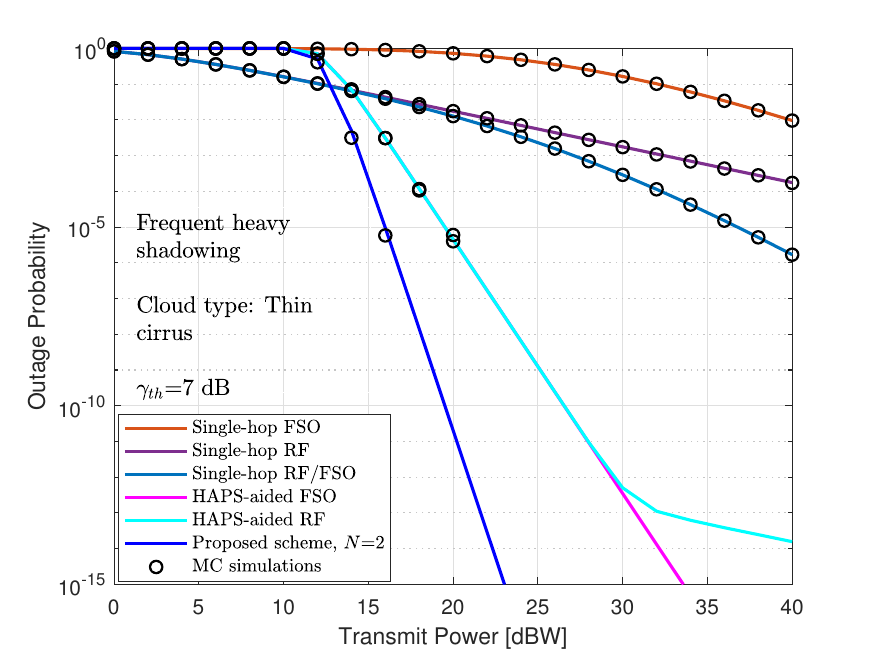}
 \caption{ {Outage probability performance of different system models for downlink SatCom under clear weather conditions.}}
  \label{fig1}
\end{figure}

{In Fig. \ref{fig1}, we have compared the outage performance of the proposed scheme with the single-hop FSO, single-hop RF, single-hop hybrid RF/FSO, HAPS-aided FSO, and HAPS-aided RF systems with respect to the transmit power and assuming the same atmospheric conditions. It is clear from the plots that the outage performance of the proposed scheme is better than all other systems. 
In addition, the figure shows that single-hop hybrid RF/FSO performs better than single-hop RF and single-hop FSO and this gain is obtained from the hybrid communication. It is also inferred from the figure that  using a HAPS as the relay node improves the overall communication. This is due to the fact that the FSO link from the satellite to the HAPS node is less vulnerable to atmospheric attenuation. Also, for single-hop RF communication, the simulation results have shown that the outage performance is highly degraded by oxygen attenuation due to the large distance between the satellite and the GS. Furthermore, we observe a validation of the theoretical results with the MC simulations, which justifies the correctness of our derivations.} Finally, it is clear from the figure that the outage probability decreases when the transmit power increases.

\begin{figure}[!t]
  \centering
    \includegraphics[width=3.6in]{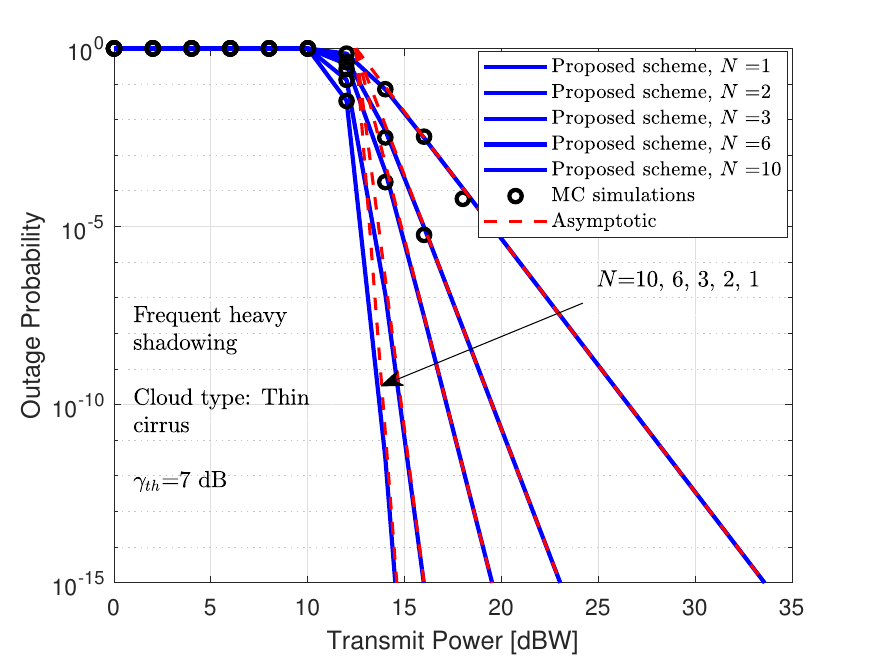}
  \caption{ {Impact of the HAPS selection scheme on the proposed model in terms of outage probability.}}
  \label{figSelc}
\end{figure}

Fig. \ref{figSelc} depicts the outage probability performance for several HAPS nodes in clear weather conditions. As we can see from the figure, increasing $N$ improves the overall performance. {At an outage of $10^{-5}$, we can observe a gain of $5$ dB between the curves of $N=10$ and $N=1$. Thus, the proposed HAPS selection scheme significantly improves the dual-hop HAPS-aided communication.} Furthermore, the theoretical results are validated with the MC simulations {for different number of HAPS systems. In addition, the figure shows that the asymptotic outage probability curves almost match the exact outage probability curves for the high SNR region, which validate the obtained derivations.}

\subsection{Impact of Aperture Averaging}
\begin{figure}[!t]
  \centering
    \includegraphics[width=3.6in]{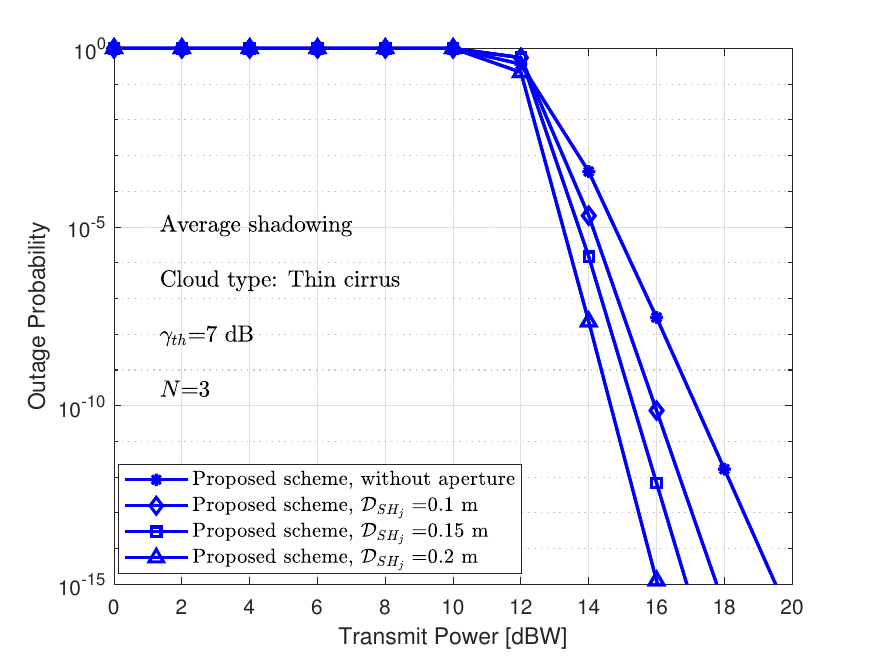}
    \caption{Outage probability performance of the proposed scheme for various aperture sizes.}
  \label{fig:aperture}
\end{figure}
\begin{figure}[!t]
  \centering
    \includegraphics[width=3.6in]{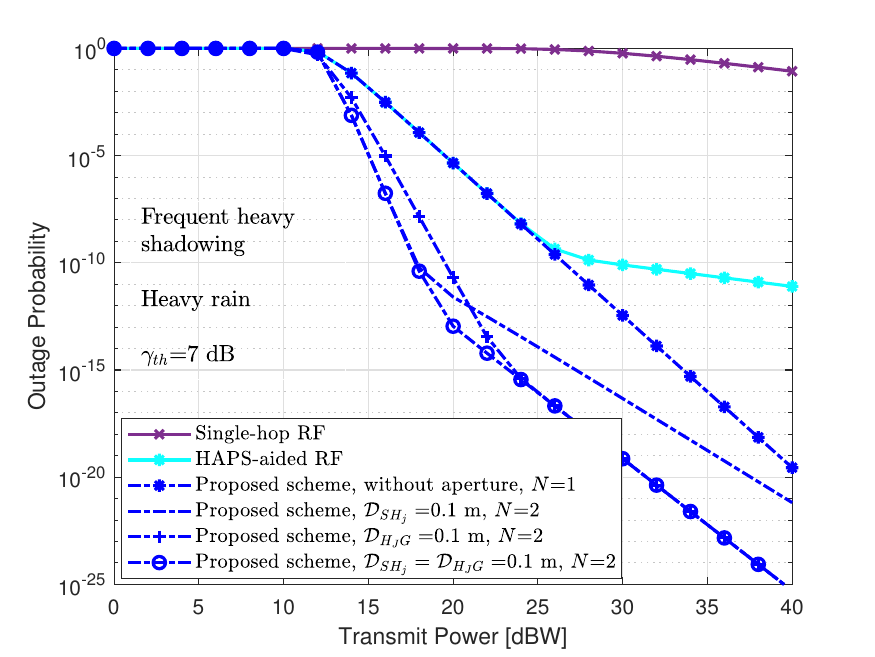}
    \caption{{Outage probability performance of the proposed scheme for various aperture sizes under heavy rain weather.}}
  \label{fig:heavyrain}
\end{figure}

{In this subsection, we analyze the impact of the aperture averaging technique on the proposed model in terms of outage probability. \\
In Fig. \ref{fig:aperture}, we compared the performance of the proposed setup for different aperture sizes for $S$ to $H_j$ communication under clear weather conditions for $N=3$ HAPS nodes.}
As the figure indicates, the use of greater aperture sizes increases the gain in terms of transmit power and improves the overall performance by reducing the effect of turbulence-induced fading. This is because, with an increase in the aperture size, more energy is collected by the receiver beam increases and thus offers more power gain. {Similarly, Fig. \ref{fig:heavyrain} compares the outage performance of the proposed model for different aperture sizes in the presence of heavy rain weather. As we can see from the figure, the single-hop RF system is highly degraded by heavy rain conditions and a significant improvement is observed with the use of HAPS node. Also, it is noticed that hybrid RF/FSO without aperture averaging and with $N=1$ shows better performance than HAPS-aided RF. Thus, the use of FSO backup link helps in enhancing the outage performance in the presence of heavy rain. For our proposed model, we compared the use of the aperture averaging technique when it is only considered for $S$ to $H_j$ communication or for $H_J$ to $G$ communication, and at both hops. It is inferred from the figure that considering aperture for $S$ to $H_j$ communication improves the outage performance only at low transmit power, whereas, assuming the aperture averaging for $H_J$ to $G$ communication shows better performance at high transmit power. Also, as expected, using the aperture averaging technique at both hops highly increases the performance gain.}
{\subsection{Impact of Weather Conditions}}
\begin{figure}[!t]
  \centering
    \includegraphics[width=3.6in]{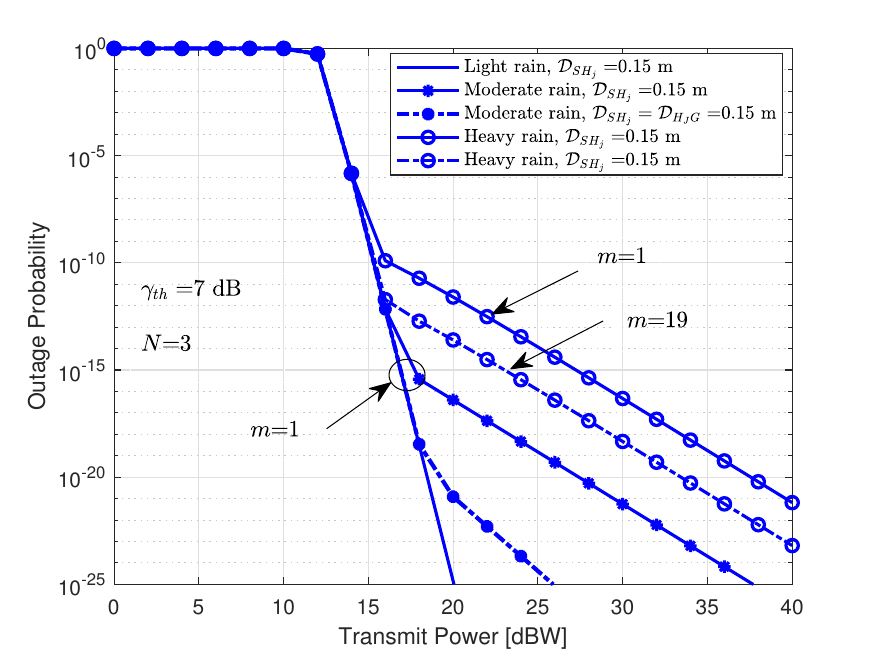}
     \caption{Outage probability performance of the proposed scheme under rainy weather conditions.}
  \label{fig:rain}
\end{figure}
\begin{figure}[!t]
  \centering
    \includegraphics[width=3.6in]{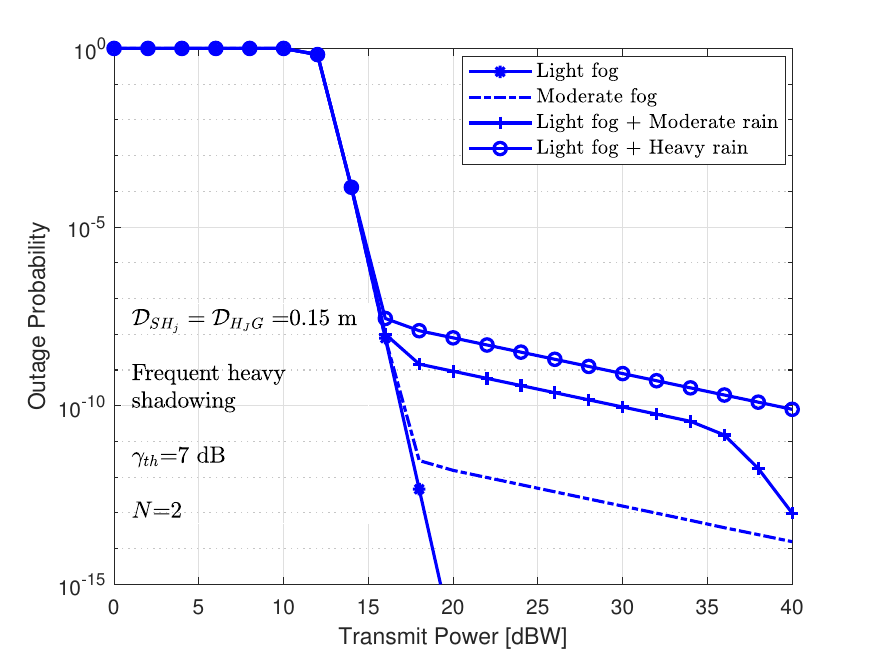}
     \caption{Outage probability performance of the proposed model under rainy and foggy weather conditions.}
  \label{fig:fog}
\end{figure}
In Fig. \ref{fig:rain}, we observe the outage performance of the proposed scheme for different rain levels for $N=3$, while considering the aperture averaging technique. {As expected, increasing the rain rate, deteriorates the overall performance} as the attenuation level increases. Also, we can see that decreasing the severity of fading for the RF link, improves the outage performance for heavy rain. {Moreover, the figure shows better performance when using the aperture averaging technique at both hops for moderate rain state.}
Finally, the simulation results show that the RF communication is highly affected by rain and that the hybrid communication relies on the FSO link under rainy conditions. 

Fig. \ref{fig:fog} shows the impact of foggy and rainy weather on the performance of the system for $N=2$ and considering an aperture size of {$\mathcal{D}_{SH_j}=\mathcal{D}_{H_JG}=0.15$ m}. We first consider light fog, which can be present up to 100 m above ground level. Then, increasing the thickness of the fog layer from light to moderate deteriorates the performance of the FSO communication as shown in the figure, and the RF link becomes dominant for hybrid communication{s}. Furthermore, in the presence of light fog with all rain levels, the overall performance is degraded as both links are affected, however, the use of the aperture averaging at both hops, helps to mitigate these effects.

\subsection{Impact of Pointing Errors}
\begin{figure}[!t]
  \centering
    \includegraphics[width=3.6in]{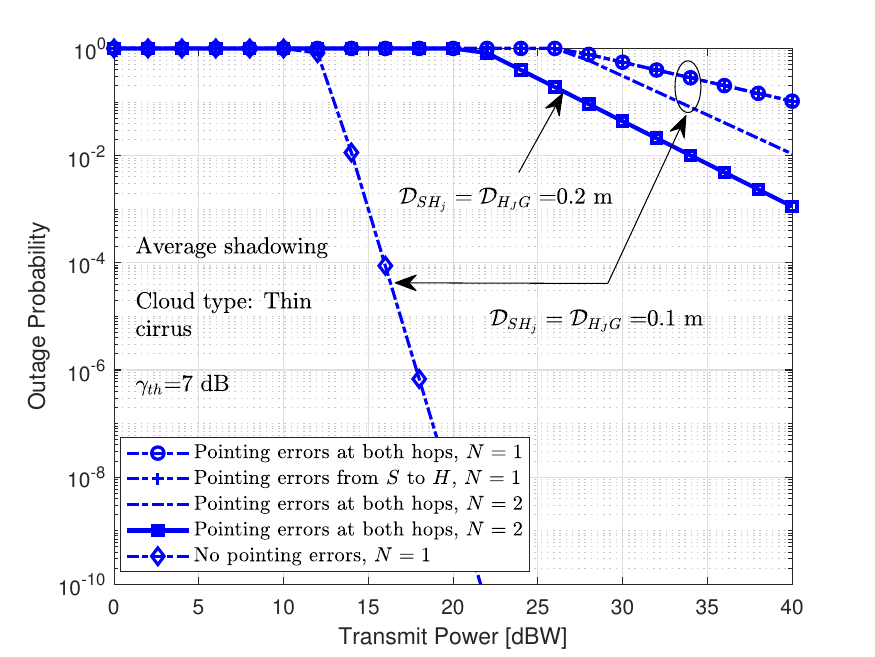}
    \caption{The impact of pointing errors on the proposed model in terms of outage probability.}
  \label{fig:PE}
\end{figure}
 
{In Fig. \ref{fig:PE}, we study the impact of zero boresight pointing errors on the proposed model. It is observed from the figure that that severe deterioration occurs in the outage performance due to pointing errors phenomenon. In addition, the simulation results reveal that no improvement is noticed in the presence of pointing errors only for $S$ to $H_j$ communication, and this is due to the fact that the link from $S$ to $H_j$ suffers from serious deterioration. Thus, it can be noted that the overall performance is degraded because of the misalignment between the transmitter and receiver. However, we can see that increasing the aperture size and employing HAPS selection can help us to alleviate this deterioration.}

\subsection{Impact of Wind Speed}
\begin{figure}[!t]
  \centering
    \includegraphics[width=3.6in]{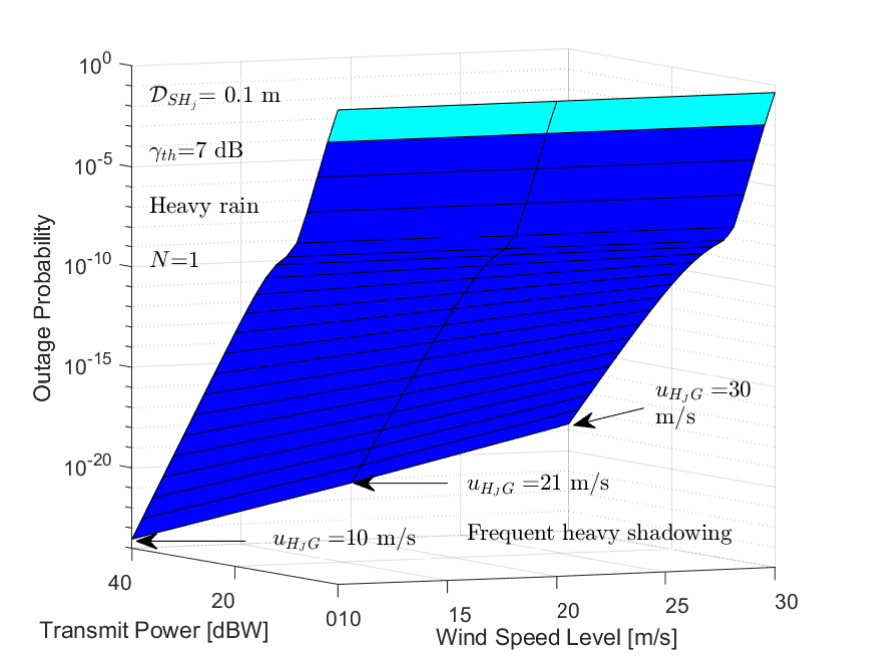}
    \caption{Outage probability performance of the proposed scheme for different wind speeds.}
  \label{fig:wind}
\end{figure}

We also investigate the outage performance for different wind speed levels at the GS {for $N$=1 and for $\mathcal{D}_{SH_j}$=0.1 m}. For low, moderate, and strong wind speed, $u_{H_JG}$ is set to $u_{H_JG}$ = 10 m/s, $u_{H_JG}$ = 21 m/s, and $u_{H_JG}$ = 30 m/s, respectively for the FSO link between HAPS and GS. As we see in Fig. \ref{fig:wind}, increasing {the wind velocity deteriorates the overall outage performance. This is due to the fact that increasing the wind speed level leads to a displacement of the beams, and as the wind speed is directly related to the scintillation index, it causes greater atmospheric turbulence.} Furthermore, at a lower wind speed $u_{H_JG}$ = 10 m/s, we can see a significant power gain compared to $u_{H_JG}$ = 30 m/s.

\subsection{Design Guidelines}
In this subsection, we provide guidelines that can helpful for the design of HAPS-aided SatCom downlink systems.
\begin{itemize}
\item{ The proposed hybrid RF/FSO model shows better performance than single-hop FSO and RF, single-hop hybrid RF/FSO, HAPS-aided RF, and HAPS-aided FSO in terms of outage probability in clear weather conditions.}
\item The use of HAPS improves SatCom's performance as the link from satellite to HAPS is less affected by the atmospheric turbulence and attenuation.
  \item The simulations have shown that the FSO channel is slightly affected by rainy weather, whereas the RF link is highly affected, although it remains available. Moreover, the presence of foggy weather deteriorates the FSO communication.
    \item The zenith angle is directly related to the performance of downlink SatCom. In fact, for lower zenith angle values, we observe lower atmospheric attenuation and this can enhance the overall performance.
    \item Aperture averaging should be considered as it can mitigate the effect of turbulence-induced fading and improve performance, especially for higher aperture diameter values. {Also, it was inferred from the simulation results that using aperture averaging at the HAPS node improves the outage probability at low transmit power, whereas using aperture averaging at the GS station shows enhanced performance at high transmit power. }
    \item The misalignment between the transmitter and the receiver caused by pointing errors substantially degrades the overall {outage} performance. 
 \item Significant performance improvement in terms of power gain is obtained, in the presence of misalignment, by increasing the aperture averaging size.
    \item The HAPS selection based on the satellite-HAPS channel quality improves the overall performance.
\end{itemize}

\section{Conclusion}
In this paper, we proposed a new HAPS-assisted downlink SatCom model with hybrid RF/FSO communication. More precisely, in the first phase of transmission, the best HAPS node was selected among multiple HAPS nodes, and then in the second phase, we focused on the simultaneous transmission on both RF and FSO links. For the proposed model, the outage probability expressions were derived, {along with outage probability analysis at high SNR,} and MC simulations were provided to validate the accuracy of our analytical results. Furthermore, we considered different weather conditions and investigated the impact of pointing errors, temperature, aperture averaging technique, and wind speed. The simulations indicated that zero-boresight pointing errors lead to severe performance impairments and that the aperture averaging can mitigate the effects of turbulence-induced fading and misalignment caused by pointing errors. Moreover, the HAPS selection based on the satellite-HAPS channel was shown to enhance the overall performance. Finally, guidelines were provided for the design of a HAPS-aided SatCom system.

\balance
\bibliographystyle{IEEEtran}
\bibliography{refere}

\begin{thebibliography}{10}
\providecommand{\url}[1]{#1}
\csname url@samestyle\endcsname
\providecommand{\newblock}{\relax}
\providecommand{\bibinfo}[2]{#2}
\providecommand{\BIBentrySTDinterwordspacing}{\spaceskip=0pt\relax}
\providecommand{\BIBentryALTinterwordstretchfactor}{4}
\providecommand{\BIBentryALTinterwordspacing}{\spaceskip=\fontdimen2\font plus
\BIBentryALTinterwordstretchfactor\fontdimen3\font minus
  \fontdimen4\font\relax}
\providecommand{\BIBforeignlanguage}[2]{{%
\expandafter\ifx\csname l@#1\endcsname\relax
\typeout{** WARNING: IEEEtran.bst: No hyphenation pattern has been}%
\typeout{** loaded for the language `#1'. Using the pattern for}%
\typeout{** the default language instead.}%
\else
\language=\csname l@#1\endcsname
\fi
#2}}
\providecommand{\BIBdecl}{\relax}
\BIBdecl

\bibitem{alam2020}
M.~S. {Alam}, G.~K. {Kurt}, H.~{Yanikomeroglu}, P.~{Zhu}, and N.~D. {D}{\`a}o,
  ``High altitude platform station based super macro base station
  constellations,'' \emph{IEEE Commun. Magazine}, vol.~7, no.~1, pp. 103--109,
  2021.

\bibitem{guo2020p}
K.~Guo, M.~Lin, B.~Zhang, J.-B. Wang, Y.~Wu, W.-P. Zhu, and J.~Cheng,
  ``Performance analysis of hybrid satellite-terrestrial cooperative networks
  with relay selection,'' \emph{IEEE Trans. Veh. Technol.}, vol.~69, no.~8, pp.
  9053--9067, 2020.

\bibitem{ITU2016}
ITU, ``Radio regulations articles.''\hskip 1em plus 0.5em minus 0.4em\relax
  International Telecommunication Union, Recommendation, 2016.

\bibitem{9380673}
G.~Karabulut~Kurt, M.~G. Khoshkholgh, S.~Alfattani, A.~Ibrahim, T.~S.~J.
  Darwish, M.~S. Alam, H.~Yanikomeroglu, and A.~Yongacoglu, ``A vision and
  framework for the high altitude platform station ({HAPS}) networks of the
  future,'' \emph{IEEE Commun. Surveys Tuts.}, vol.~23, no.~2, pp. 729--779,
  2021.

\bibitem{9530144}
N.~Vishwakarma and S.~R, ``Capacity analysis of adaptive combining for hybrid
  {FSO/RF} satellite communication system,'' in \emph{National Conference on
  Communications (NCC)}, 2021, pp. 1--6.

\bibitem{2020review}
S.~C. Arum, D.~Grace, and P.~D. Mitchell, ``A review of wireless communication
  using high-altitude platforms for extended coverage and capacity,''
  \emph{Computer Communications}, vol. 157, pp. 232--256, 2020.

\bibitem{andrews2005}
L.~C. Andrews and R.~L. Phillips, ``Laser beam propagation through random media
  ({SPIE} {P}ress {M}onograph).''\hskip 1em plus 0.5em minus 0.4em\relax
  Bellingham, WA, USA: SPIE, 2005.

\bibitem{weather}
F.~Nadeem, V.~Kvicera, M.~S. Awan, E.~Leitgeb, S.~S. Muhammad, and G.~Kandus,
  ``Weather effects on hybrid {FSO/RF} communication link,'' \emph{IEEE J. Sel.
  Areas Commun.}, vol.~27, no.~9, pp. 1687--1697, 2009.

\bibitem{kaushal2016optical}
H.~Kaushal and G.~Kaddoum, ``Optical communication in space: Challenges and
  mitigation techniques,'' \emph{IEEE Commun. Surveys Tuts.}, vol.~19, no.~1,
  pp. 57--96, 2016.

\bibitem{barrios2012}
R.~Barrios and F.~Dios, ``Exponentiated {W}eibull distribution family under
  aperture averaging for gaussian beam waves,'' \emph{Optics Express}, vol.~20,
  no.~12, pp. 13\,055--13\,064, 2012.

\bibitem{erdogan2020}
E.~Erdogan, I.~Altunbas, G.~K. Kurt, M.~Bellemare, G.~Lamontagne, and
  H.~Yanikomeroglu, ``Site diversity in downlink optical satellite networks
  through ground station selection,'' \emph{IEEE Access}, vol.~9, pp.
  31\,179--31\,190, 2021.

\bibitem{kazemi}
H.~Kazemi, M.~Uysal, and F.~Touati, ``Outage analysis of hybrid {FSO/RF}
  systems based on finite-state {M}arkov chain modeling,'' \emph{in Int.
  Workshop in Optical Wireless Commun. (IWOW)}, pp. 11--15, 2014.

\bibitem{krishnan}
P.~Krishnan, ``Performance analysis of hybrid {RF/FSO} system using {BPSK-SIM}
  and {DPSK-SIM} over {G}amma-{G}amma turbulence channel with pointing errors
  for smart city applications,'' \emph{IEEE Access}, vol.~6, pp.
  75\,025--75\,032, 2018.

\bibitem{bag2018}
B.~Bag, A.~Das, I.~S. Ansari, A.~Proke{\v{s}}, C.~Bose, and A.~Chandra,
  ``Performance analysis of hybrid {FSO} systems using {FSO/RF-FSO} link
  adaptation,'' \emph{IEEE Photon. J.}, vol.~10, no.~3, pp. 1--17, 2018.

\bibitem{bag2019}
B.~Bag, A.~Das, C.~Bose, and A.~Chandra, ``Hybrid {FSO/RF-FSO} systems over
  generalized {M}{\'a}laga distributed channels with pointing errors,'' in
  \emph{European Signal Process. Conf. (EUSIPCO)}, 2019, pp. 1--5.

\bibitem{amirabadi}
M.~A. Amirabadi and V.~T. Vakili, ``A novel hybrid {FSO/RF} communication
  system with receive diversity,'' \emph{Optik}, vol. 184, pp. 293--298, 2019.

\bibitem{2018selection}
V.~Sundharam and S.~Johari, ``Selection {C}ombining in hybrid {RF/FSO} systems
  for {IM/DD} and heterodyne detection in varying weather conditions,'' in
  \emph{Micro-Electronics and Telecom Eng (ICMETE)}, 2018, pp. 286--291.

\bibitem{ma2015performance}
J.~Ma, K.~Li, L.~Tan, S.~Yu, and Y.~Cao, ``Performance analysis of
  satellite-to-ground downlink coherent optical communications with spatial
  diversity over {G}amma--{G}amma atmospheric turbulence,'' \emph{Applied
  optics}, vol.~54, no.~25, pp. 7575--7585, 2015.

\bibitem{swaminathan}
R.~Swaminathan, S.~Sharma, and A.~MadhuKumar, ``Performance analysis of
  {HAPS}-based relaying for hybrid {FSO/RF} downlink satellite communication,''
  in \emph{Veh. Technol. Conf. (VTC2020-Spring)}.\hskip 1em plus 0.5em minus
  0.4em\relax IEEE, 2020, pp. 1--5.

\bibitem{swamina}
R.~Swaminathan, S.~Sharma, N.~Vishwakarma, and A.~Madhukumar, ``{HAPS}-based
  relaying for integrated space-air-ground networks with hybrid {FSO/RF}
  communication: A performance analysis,'' \emph{IEEE Trans. Aerosp. Electron.
  Syst.}, vol.~17, pp. 1--17, 2021.

\bibitem{9446153}
S.~Shah, M.~Siddharth, N.~Vishwakarma, R.~Swaminathan, and A.~S. Madhukumar,
  ``Adaptive-combining-based hybrid {FSO/RF} satellite communication with and
  without {HAPS},'' \emph{IEEE Access}, vol.~9, pp. 81\,492--81\,511, 2021.

\bibitem{barrios2013}
R.~Barrios~Porras, ``Exponentiated {W}eibull fading channel model in free-space
  optical communications under atmospheric turbulence.''\hskip 1em plus 0.5em
  minus 0.4em\relax Ph.D. dissertation, Dept. Signal Theory Commun., Univ.
  Polit{\`e}cnica de Catalunya ({UPC}), {B}arcelona, {S}pain, May 2013.

\bibitem{ITUR}
ITU, ``Propagation data required for the design of {E}arth-space systems
  operating between 20 {TH}z and 375 {TH}z.''\hskip 1em plus 0.5em minus
  0.4em\relax International Telecommunication Union, Recommendation P.1622,
  2003.

\bibitem{touati}
A.~Touati, A.~Abdaoui, F.~Touati, M.~Uysal, and A.~Bouallegue, ``On the effects
  of combined atmospheric fading and misalignment on the hybrid {FSO/RF}
  transmission,'' \emph{J. of Optical Commun. and Netw.}, vol.~8, no.~10, pp.
  715--725, 2016.

\bibitem{ITUrain}
ITU, ``Specific attenuation model for rain for use in prediction
  methods.''\hskip 1em plus 0.5em minus 0.4em\relax International
  Telecommunication Union, Recommendation P.838-3, 2003.

\bibitem{2019physical}
Y.~Ai, A.~Mathur, M.~Cheffena, M.~R. Bhatnagar, and H.~Lei, ``Physical layer
  security of hybrid satellite-{FSO} cooperative systems,'' \emph{IEEE Photon.
  J.}, vol.~11, no.~1, pp. 1--14, 2019.

\bibitem{aragon2008}
A.~Aragon-Zavala, J.~L. Cuevas-Ru{\'\i}z, and J.~A. Delgado-Pen{\'\i}n,
  \emph{High-{A}ltitude {P}latforms for {W}ireless {C}ommunications}.\hskip 1em
  plus 0.5em minus 0.4em\relax Wiley Online Library, 2008, vol.~5.

\bibitem{2010optical}
F.~Fidler, M.~Knapek, J.~Horwath, and W.~R. Leeb, ``Optical communications for
  high-altitude platforms,'' \emph{IEEE J. Sel. Topics Quantum Electron.},
  vol.~16, no.~5, pp. 1058--1070, 2010.

\bibitem{giggenbach}
D.~Giggenbach, R.~Purvinskis, M.~Werner, and M.~Holzbock, ``Stratospheric
  optical inter-platform links for high altitude platforms,'' in \emph{Int.
  Commun. Satellite Systems Conf. and Exhibit}, 2002, p. 1910.

\bibitem{2018performance}
Y.~Wang, P.~Wang, X.~Liu, and T.~Cao, ``On the performance of dual-hop mixed
  {RF/FSO} wireless communication system in urban area over aggregated
  exponentiated {W}eibull fading channels with pointing errors,'' \emph{Optics
  Commun.}, vol. 410, pp. 609--616, 2018.

\bibitem{yang2014free}
F.~Yang, J.~Cheng, and T.~A. Tsiftsis, ``Free-space optical communication with
  nonzero boresight pointing errors,'' \emph{IEEE Trans. on Commun.}, vol.~62,
  no.~2, pp. 713--725, 2014.

\bibitem{ITUR2003}
ITU, ``Prediction methods required for the design of {E}arth-space systems
  operating between 20 {TH}z and 375 {TH}z.''\hskip 1em plus 0.5em minus
  0.4em\relax International Telecommunication Union, Recommendation P.1622,
  2003.

\bibitem{awan2009}
M.~S. Awan, E.~Leitgeb, B.~Hillbrand, F.~Nadeem, M.~Khan \emph{et~al.}, ``Cloud
  attenuations for free-space optical links,'' in \emph{Int. Workshop on
  Satellite and Space Commun.}\hskip 1em plus 0.5em minus 0.4em\relax IEEE,
  2009, pp. 274--278.

\bibitem{johari2017}
S.~Johari and V.~Sundharam, ``Performance analysis of {IM/DD} vs. heterodyne
  detection techniques of an {E}arth-satellite {FSO} link for next generation
  wireless communication,'' in \emph{IEEE Malaysia Int. Conf. on Commun.
  (MICC)}, 2017, pp. 191--196.

\bibitem{2018outage}
E.~T. Michailidis, N.~Nomikos, P.~Bithas, D.~Vouyioukas, and A.~G. Kanatas,
  ``Outage probability of triple-hop mixed {RF/FSO/RF} stratospheric
  communication systems,'' in \emph{IEEE Int. Conf. on Adv. in Satellite and
  Space Commun. (SPACOMM)}, 2018, pp. 1--6.

\bibitem{Wolform}
\BIBentryALTinterwordspacing
The {W}olfram functions site. [Online]. Available: \url{http://www.wolfram.com}
\BIBentrySTDinterwordspacing

\bibitem{2014table}
I.~S. Gradshteyn and I.~M. Ryzhik, \emph{Table of Integrals, Series, and
  Products}.\hskip 1em plus 0.5em minus 0.4em\relax Academic Press, 2014.

\bibitem{el2011trends}
A.~El~Oualkadi, \emph{Trends and {C}hallenges in {CMOS} {D}esign for {E}merging
  60 {GH}z {WPAN} {A}pplications}.\hskip 1em plus 0.5em minus 0.4em\relax
  InTech, 2011.

\end{thebibliography}
\end{document}